\documentclass{PoS}
\usepackage{amsmath}
\usepackage{amssymb}
\usepackage{wrapfig}
\usepackage{xspace}
\usepackage{cite}
\usepackage{setspace}
\usepackage[square,numbers,sort&compress]{natbib}

\title{Squark/gluino searches in hadronic channels with CMS}

\ShortTitle{Squark/gluino searches in hadronic channels with CMS}

\author{Tai Sakuma for the CMS collaboration\\
        University of Bristol\\
        E-mail: \email{tai.sakuma@cern.ch}}

\abstract{These proceedings summarize the results of four analyses
  which searched for squarks and gluinos in hadronic final states with
  missing transverse momentum in 2.3~fb$^{-1}$ of data in
  proton-proton collisions at $\sqrt{s}=13$~TeV collected in the year
  2015 with the CMS detector at the CERN LHC. Each analysis is
  characterized by a different kinematic variable that is sensitive to
  the presence of invisible particles, e.g., $M_\text{T2}$,
  $\alpha_{\text{T}}$, and \textit{razor} variables. We observed no
  significant deviation from the standard model prediction and placed
  limits on the production cross sections and the masses of squarks
  and gluinos in simplified models of supersymmetric models. The
  limits are significantly extended from the previous results.

}

\FullConference{Fourth Annual Large Hadron Collider Physics\\
		13-18 June 2016\\
		Lund, Sweden}

\begin{document}

\graphicspath{{images/}}
\DeclareGraphicsExtensions{.pdf,.png}

\newcommand{\GeV}{\ensuremath{\,\text{Ge\hspace{-.08em}V}}\xspace}
\newcommand{\TeV}{\ensuremath{\,\text{Te\hspace{-.08em}V}}\xspace}

\newcommand{\cPgg}{\ensuremath{\gamma}} 
\providecommand{\PGm}{\ensuremath{\mu}\xspace} 

\providecommand{\PSg}{\ensuremath{\widetilde{\cmsSymbolFace{g}}}\xspace} 
\providecommand{\PSQ}{\ensuremath{\widetilde{\cmsSymbolFace{q}}}\xspace} 
\providecommand{\PSQt}{\ensuremath{\widetilde{\cmsSymbolFace{t}}}\xspace} 
\providecommand{\PSQb}{\ensuremath{\widetilde{\cmsSymbolFace{b}}}\xspace}
\providecommand{\PSGcz}{\ensuremath{\widetilde{\chi}^0}\xspace} 

\newcommand{\pt}{\ensuremath{p_{\mathrm{T}}}\xspace}
\newcommand{\ET}{\ensuremath{E_{\mathrm{T}}}\xspace}
\newcommand{\HT}{\ensuremath{H_{\mathrm{T}}}\xspace}
\newcommand{\ETm}{\ensuremath{E_{\mathrm{T}}^{\text{miss}}}\xspace}
\newcommand{\MET}{\ETm}
\newcommand{\VEtmiss}{\ensuremath{{\vec E}_{\mathrm{T}}^{\text{miss}}}\xspace}
\newcommand{\ptvec}{\ensuremath{{\vec p}_{\mathrm{T}}}\xspace}
\newcommand{\ptvecmiss}{\ensuremath{{\vec p}_{\mathrm{T}}^{\kern1pt\text{miss}}}\xspace}

\newcommand{\cmsSymbolFace}{\mathrm}
\newcommand{\ttbar}{\ensuremath{\cmsSymbolFace{t}\overline{\cmsSymbolFace{t}}}\xspace} 
\newcommand{\cPgn}{\ensuremath{\nu}} 
\newcommand{\cPagn}{\ensuremath{\overline{\nu}}} 
\newcommand{\cPZ}{\ensuremath{\cmsSymbolFace{Z}}} 

\providecommand{\Pe}{\ensuremath{\cmsSymbolFace{e}}\xspace}
\newcommand{\HTm}{\ensuremath{H_{\mathrm{T}}^{\text{miss}}}\xspace}
\newcommand{\MHT}{\HTm}
\newcommand{\HTvecmiss}{\ensuremath{{\vec H}_{\mathrm{T}}^{\kern1pt\text{miss}}}\xspace}

\newcommand{\njet}{\ensuremath{n_{\text{jet}}}\xspace}
\newcommand{\nb}{\ensuremath{n_{\text{b}}}\xspace}

\newcommand{\alphat}{\ensuremath{\alpha_{\text{T}}}\xspace}
\newcommand{\bdphi}{\ensuremath{\Delta\phi^{\ast}}\xspace}
\newcommand{\bdphimin}{\ensuremath{\bdphi_\text{min}}\xspace}
\newcommand{\dphi}{\ensuremath{\Delta\phi}\xspace}

\newcommand{\MTtwo}{\ensuremath{M_\text{T2}}\xspace}

\newcommand{\zll}{\ensuremath{\cPZ\to \ell^{+}\ell^{-}}\xspace}
\newcommand{\gjets}{{{\cPgg}+jets}\xspace}
\newcommand{\znn}{\ensuremath{\cPZ \to \cPgn \cPagn}\xspace}
\newcommand\zlljets{{\ensuremath{\cPZ (\to \ell^{+} \ell^{-} )}+jets}\xspace}
\newcommand\znnjets{{\ensuremath{\cPZ(\to\cPgn \cPagn )}+jets}\xspace}

\providecommand{\Ptau}{\ensuremath{\tau}\xspace}
\providecommand{\PW}{\ensuremath{\cmsSymbolFace{W}}\xspace}
\newcommand\wjets{{{\PW}+jets}\xspace}

\section{Introduction}

The standard model (SM) of particle physics is considered to be
incomplete. A main reason is its \textit{fine-tuning}, by which the
very large value of the \textit{bare} Higgs boson mass must be
extremely precisely larger than another very large value of the
\textit{cutoff} of the calculation of higher-order corrections, which,
for example, is of the order of the GUT scale ($10^{16}\GeV$), by a
small value of the physical mass of the Higgs boson (125\GeV).
\textit{Supersymmetry} (SUSY), an extension of the space-time symmetry
which must be \textit{broken} if it is a symmetry of nature, provides
a solution to this problem. The SM can be extended to have a broken
SUSY so as to alleviate the dependency of the Higgs boson mass
correction on the cutoff from quadratic to logarithmic and reduce the
fine-tuning to a level that can be regarded as \textit{natural}.

Supersymmetric extensions of the SM predict that each particle in the
SM has a heavier partner (\textit{superpartner}) that has not been
observed. For the level of the tuning to be natural, some of the
superpartners, such as \textit{top squarks}, \textit{bottom squarks},
and \textit{gluinos}, must have masses of the order of TeV and can be
produced in proton-proton (pp) collisions at the CERN LHC. Squark
pairs and gluino pairs have larger production cross sections than
other superpartners and have jets and a large missing transverse
momentum in the final states of their decay products. Therefore,
squark and gluino searches in hadronic channels are optimum for early
searches in LHC Run~2.

These proceedings summarize the results of four analyses
\cite{Khachatryan:2016kdk,Khachatryan:2016xvy,CMS-PAS-SUS-15-004,CMS-PAS-SUS-15-005}
that searched for squarks and gluinos in hadronic final states with a
large missing transverse momentum in 2.3 fb$^{-1}$ of data in pp
collisions at $\sqrt{s}=13$ TeV collected with the CMS detector in
2015, the first year of LHC Run~2. Each analysis is characterized by a
different kinematic variable used in the analysis, e.g., \MTtwo
\cite{Lester:1999tx,Barr:2003rg}, \alphat \cite{Khachatryan:2011tk},
and \textit{razor} variables \cite{Rogan:2010kb}. We found no evidence
of SUSY and placed limits on the production cross sections and the
masses of squarks and gluinos in \textit{simplified models}
\cite{ArkaniHamed:2007fw,Alwall:2008ag,Alwall:2008va,Alves:2011sq,Alves:2011wf,Chatrchyan:2013sza}
of SUSY models.

\section{Overall search procedure}

After introducing common kinematic variables used in the searches,
this section reviews an overall search procedure of SUSY searches in
hadronic final states.

The scalar sum of the transverse momenta (\pt) of the jets, denoted by
\HT, is a measure of the energy scale of the event. The missing
transverse momentum, which is sensitive to the presence of invisible
particles and a measure of their total transverse momentum, is the
negative of the vector sum of the \ptvec of all reconstructed
particles, \VEtmiss, or its magnitude, \MET. Alternatively, \HTvecmiss
and \MHT, which are, respectively, the negative of the vector sum of
the \ptvec of the jets and its magnitude, are also used. The azimuthal
angle between a jet \ptvec and \HTvecmiss (or \VEtmiss), written as
\dphi, is used to suppress QCD multijet events. The number of the jets
in the event is denoted by \njet. The number of the jets which are
tagged as originating from the bottom quark is denoted by \nb.

We start a search by defining a \textit{search (signal) region}, a
region of the phase space in which we search for a signal as an excess
of the number of the events from the SM prediction. Search regions are
specified in terms of common variables described in the previous
paragraph, i.e., \HT, \njet, \MHT, \MET, and \dphi, or special
variables such as \MTtwo, \alphat, \bdphimin and \textit{razor}
variables, introduced in later sections. Events in the search regions
are required not to have a isolated charged lepton or photon
(\textit{vetos}). Dominant SM background processes in search regions
are \znnjets, \wjets, \ttbar, and QCD multijets events. We perform a
\textit{blind} analysis, by which we do not analyze the data in the
search region until later in the procedure.

We, then, define \textit{control regions} (CRs), for example, by
inverting vetos or as \textit{sidebands} in variables used to define
the search region. CRs are used to predict the background in the
search region from the same or similar SM processes that dominate the
CRs. For example, \textit{double-lepton} CRs, in which an event is
required to have jets and a pair of electrons or muons, are dominated
by the \zlljets events, where $\ell=\Pe$, $\PGm$. A
\textit{single-photon} CR, in which an event has jets and a single
isolated photon, is dominated by the \gjets events. These CRs are used
to predict the background from the \znnjets events.
\textit{Single-lepton} CRs, which have jets and single isolated
electrons or muons, are dominated by the \wjets and \ttbar events and
are used to predict the background from these processes. A sideband of
a search region in \dphi, which is dominated by QCD multijets events,
can be used to predict the QCD multijets background.

We define \textit{categories}, i.e., events in the signal region are
categorized, for example, in \textit{bins} of \HT, \njet, and \nb. By
using the data in CRs and simulated events in various methods, we
predict the background in each category in the search region and
validate the prediction.

After the background predictions are validated, we, finally, analyze
the data in the search region (\textit{unblinding}) and compare, in
each category, with the predictions. Unless we observe a significant
discrepancy in the comparison, we interpret the results in simplified
models and place upper limits on the production cross sections of
superpartners in the models with the CLs method
\cite{Junk:1999kv,Read:2002hq} with the asymptotic formula
\cite{Cowan:2010js} and exclusion limits on the masses of
superpartners in the models by comparing with the theoretical cross
sections.

\section{Search with \MHT}
\label{sec:ra2b}

This section summarizes the analysis published in
Ref.~\cite{Khachatryan:2016kdk}, the first publication of the CMS SUSY
searches with the Run~2 data. The analysis searches high-\HT events
with four or more jets with no isolated electron, muon, or track. It
uses a combination of two search strategies used in Run~1
\cite{Chatrchyan:2014lfa,Chatrchyan:2013wxa}.

\begin{figure}[!h]
  \begin{minipage}[t]{0.45\linewidth}
    \centering
    \includegraphics[scale=0.35]{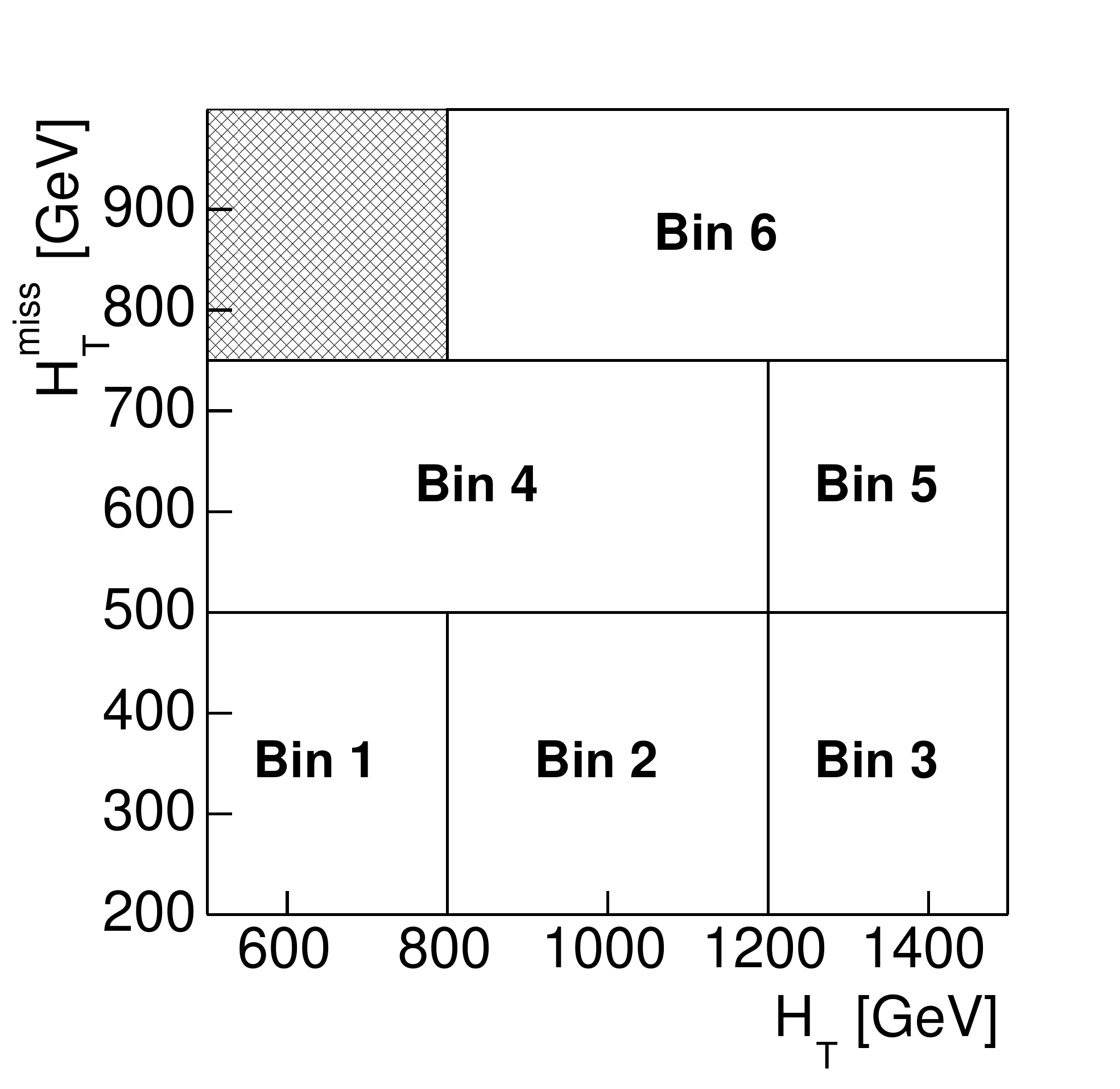}
    \caption{A schematic diagram of the categories used in
      Ref.~\cite{Khachatryan:2016kdk}, i.e., bins in \HT and \MHT.
      Events in each bin in the diagram are further binned in \njet
      and \nb.}
    \label{fig:CMS-SUS-15-002_Figure_002}
  \end{minipage}
  \hspace{0.5cm}
  \begin{minipage}[t]{0.45\linewidth}
    \centering
    \includegraphics[scale=0.35]{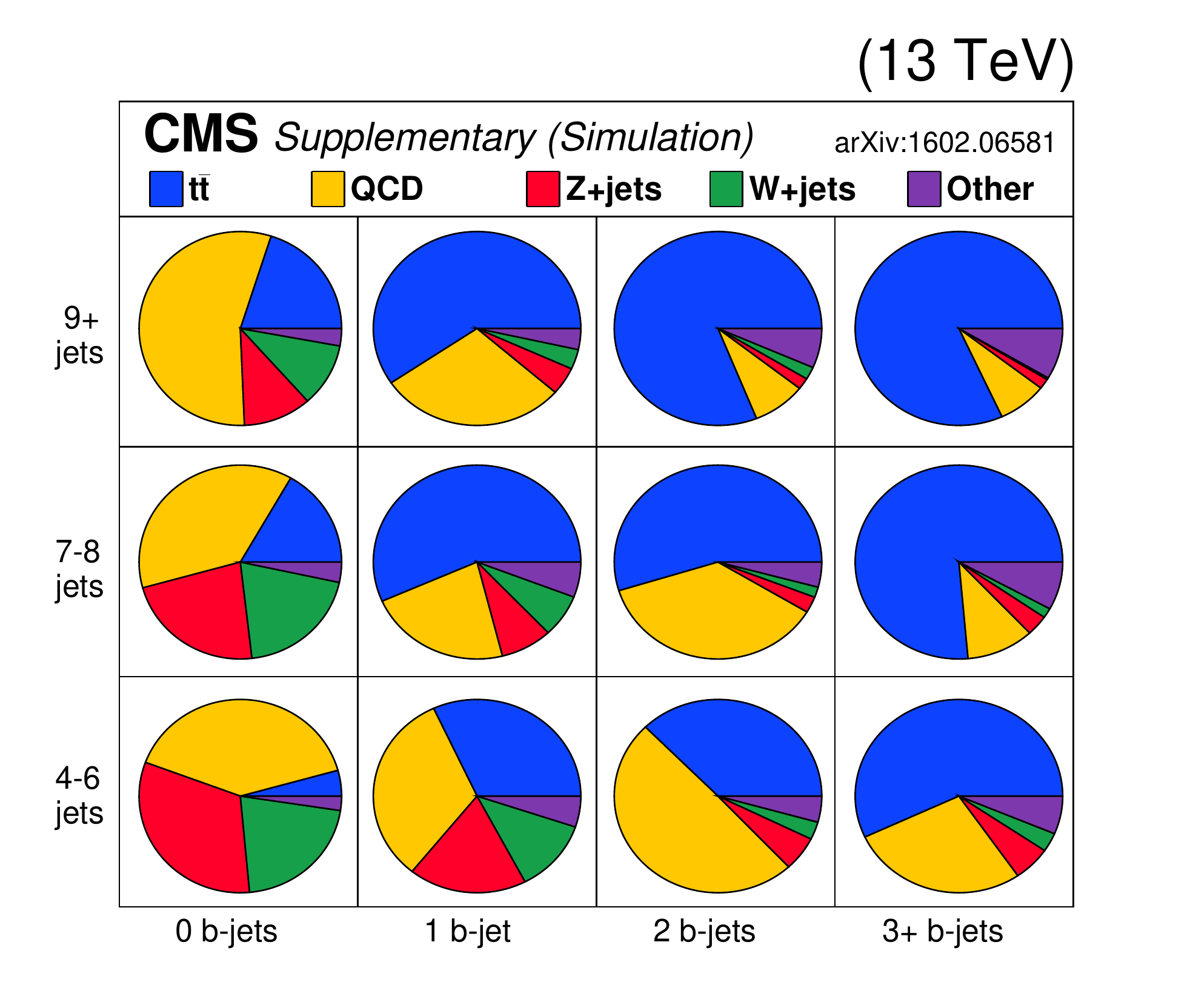}
    \caption{Background composition of the SM processes in the bins of
      \njet and \nb evaluated in simulation in the analysis in
      Ref.~\cite{Khachatryan:2016kdk}.}
    \label{fig:CMS-SUS-15-002_Figure-aux_001}
  \end{minipage}
\end{figure}

\begin{figure*}[!b]
  \centering
  \includegraphics[scale=0.5]{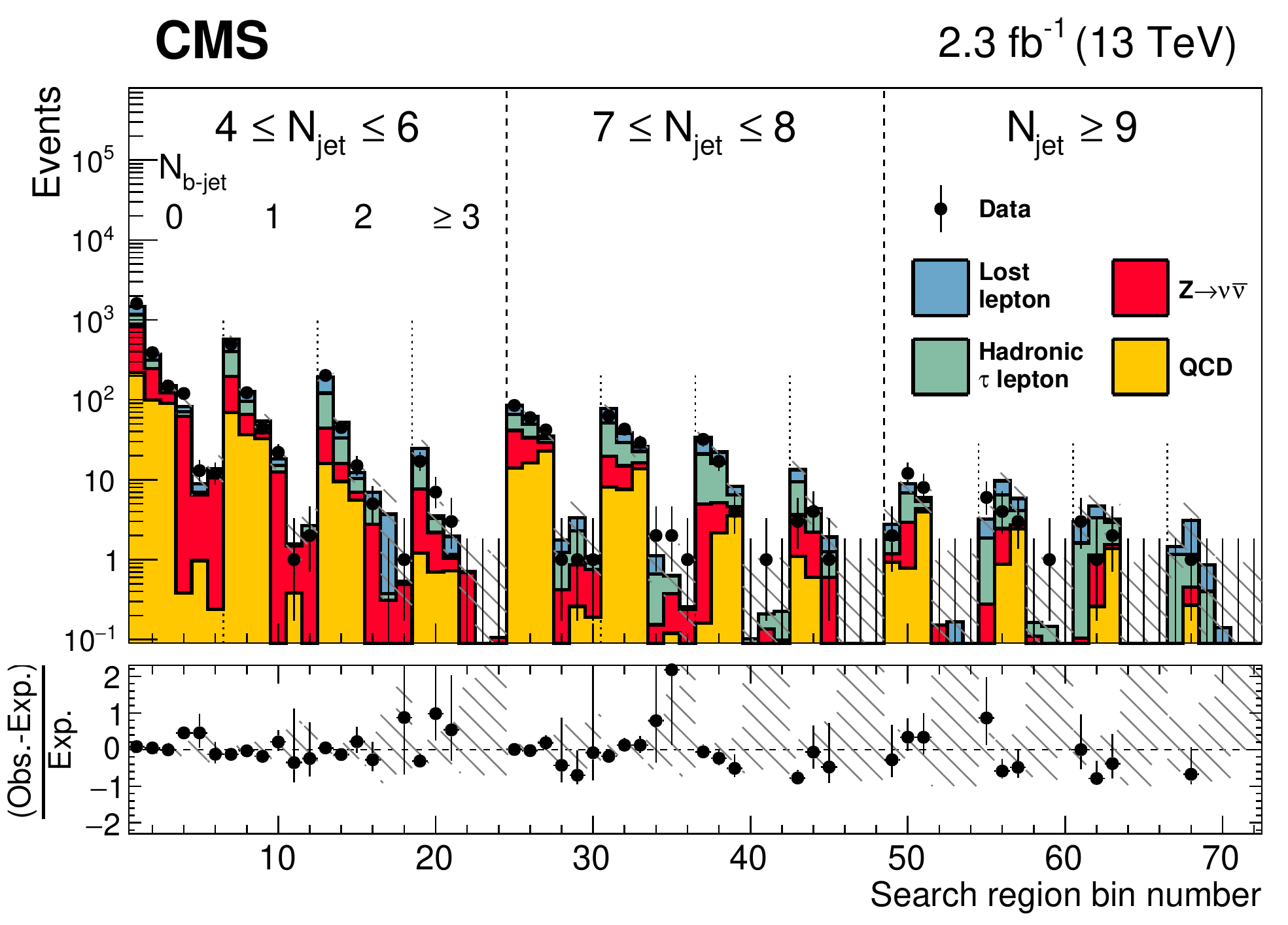}
  \caption{The observed number of the events and the SM background
    predictions in each of the 72 bins in the search region of the
    analysis in Ref.~\cite{Khachatryan:2016kdk}.}
  \label{fig:CMS-SUS-15-002_Figure_006}
\end{figure*}

The search region is specified as $\njet \ge 4$, $\HT > 500\GeV$,
$\MHT > 200\GeV$, and $\dphi > 0.5 (>~0.3)$ for the two largest \pt
jets (the 3$^{\textnormal{rd}}$ and 4$^{\textnormal{th}}$ largest \pt
jets). The events in the search region are categorized in the bins for
\HT, \MHT, \njet, and \nb. The bin boundaries in \HT and \MHT are
illustrated in Fig.~\ref{fig:CMS-SUS-15-002_Figure_002}. The intervals
of the \njet bins are 4-6, 7-8, $\ge9$, and those for the \nb bins are
0, 1, 2, $\ge3$. The analysis total has 72 bins.

The pie charts in Fig.~\ref{fig:CMS-SUS-15-002_Figure-aux_001} show
the background composition of the SM processes in the bins of \njet
and \nb evaluated in simulation. The composition noticeably changes
with \njet and \nb. For example, the lowest \njet-\nb bin contains an
approximately even mixture of the \znnjets, \wjets, and QCD multijets
events and a small fraction of the \ttbar events while the highest
\njet-\nb bin is dominated by the \ttbar events.

The background from the \wjets and \ttbar events are divided into two
groups: the \textit{lost-lepton} background, in which the electron or
muon from the \PW boson decay does not cause a veto, and the
hadronically decaying \Ptau background, in which the \Ptau lepton from
the \PW boson decay hadronically decays. The background predictions
are separately made for these two groups as well as for the \znnjets
events and the QCD multijet events. In addition to the details of the
background predictions, the evaluation of the impact of \textit{signal
  contamination} in CRs is described in
Ref.~\cite{Khachatryan:2016kdk}.

The observed number of the events and the SM background predictions in
each of the 72 bins in the search region is shown in
Fig.~\ref{fig:CMS-SUS-15-002_Figure_006}. The numerical values are
given in Ref.~\cite{Khachatryan:2016kdk}. The observed numbers are
consistent with the predictions. The interpretations of the results
are summarized in Section~\ref{sec:interpretation}.

\section{Search with \MTtwo}
\label{sec:mt2}

\begin{figure}[!b]
  \begin{minipage}[b]{0.45\linewidth}
    \centering
  \includegraphics[scale=0.35]{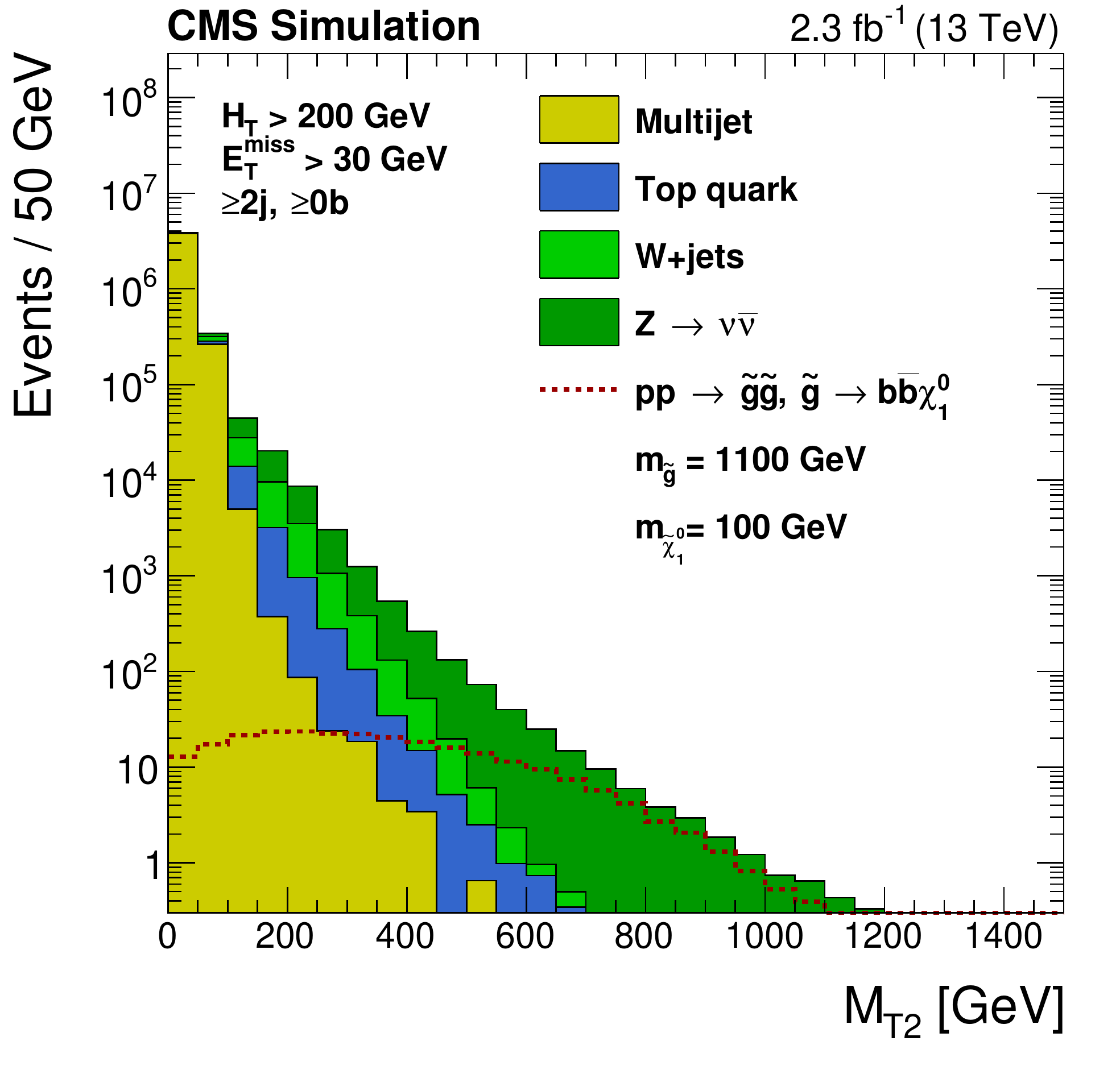}
  \caption{Distributions of \MTtwo in simulated events for the SM
    background processes and a signal process
    \cite{Khachatryan:2016xvy}.}
  \label{fig:CMS-SUS-15-003_Figure_001}
  \end{minipage}
  \hspace{0.5cm}
  \begin{minipage}[b]{0.45\linewidth}
    \centering
    \includegraphics[scale=0.30]{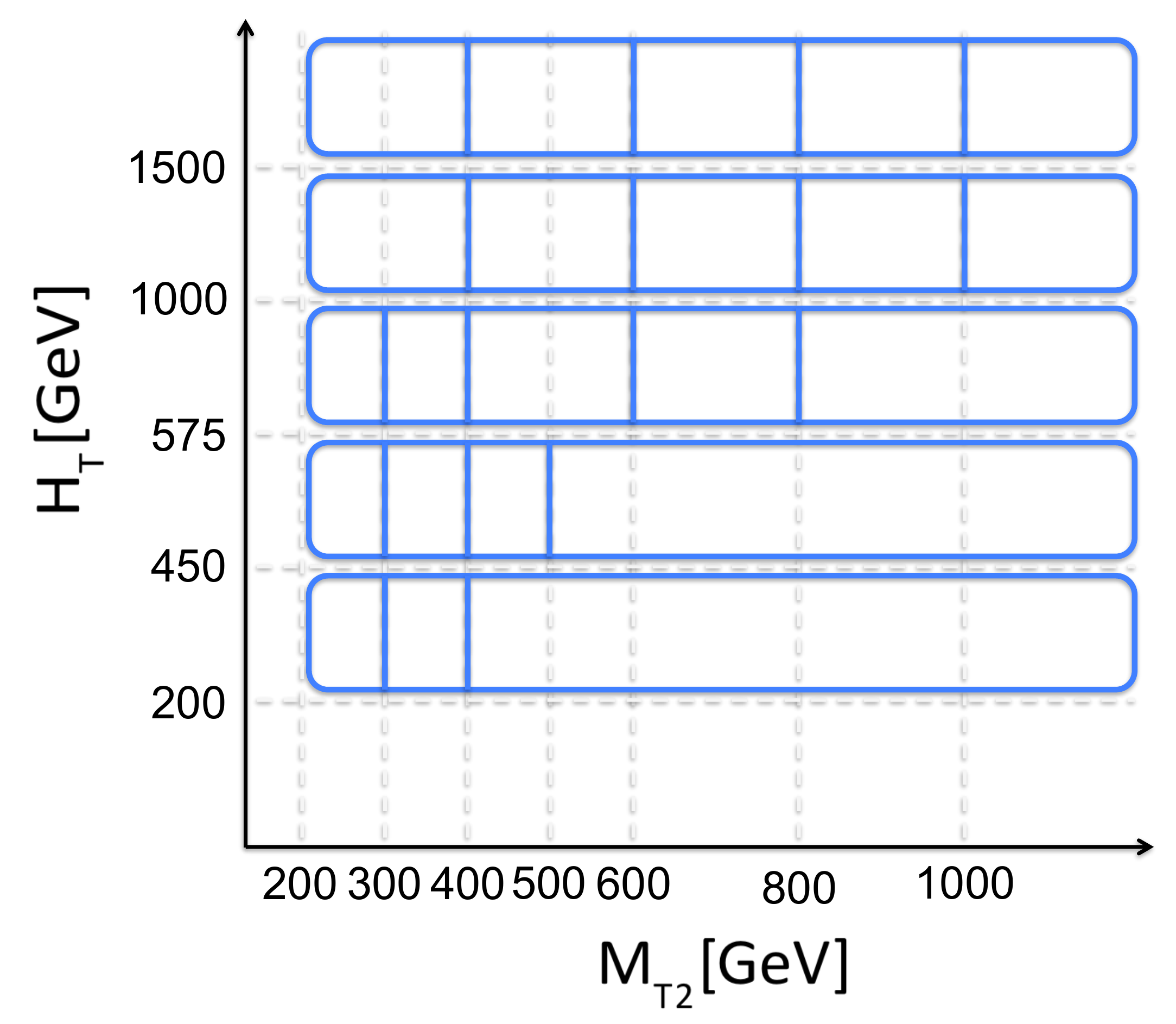}

    \vspace*{0.3cm}
    
    \includegraphics[scale=0.30]{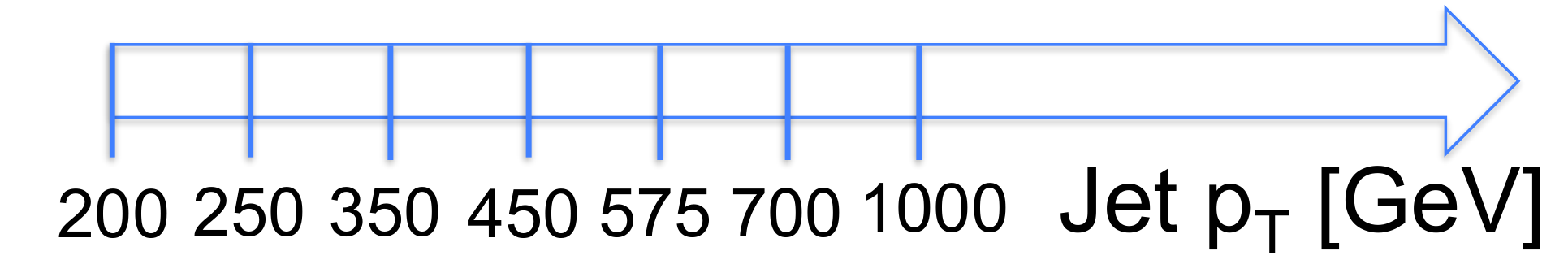}
    \caption{\textit{(top)} The boundaries of the \HT and \MTtwo bins
      for $\njet \ge 2$ and \textit{(bottom)} the boundaries of the
      jet \pt bins for $\njet = 1$ in
      Ref.~\cite{Khachatryan:2016xvy}.}
    \label{fig:CMS-SUS-15-003_Figure-aux_003-4}
  \end{minipage}
\end{figure}

In the analysis described in Ref.~\cite{Khachatryan:2016xvy}, the
\MTtwo variable is used to define the search region and categories.
This variable was also used in searches in hadronic final states in
Run~1 \cite{Khachatryan:2015vra,Chatrchyan:2012jx}.

The \MTtwo variable is an extension of the \textit{transverse mass} to
the case in which the event has two invisible particles in the final
state \cite{Lester:1999tx,Barr:2003rg}. It is defined as
\begin{eqnarray*}
  \MTtwo = \min_{\vec{q}_\textrm{T} + \vec{r}_\textrm{T} = \vec{E}_\textrm{T}^\textrm{miss}}\left[\max\left(M_\textrm{T}(\vec{p}_\textrm{T}^{\,j_1}, \vec{q}_\textrm{T}), M_\textrm{T}(\vec{p}_\textrm{T}^{\,j_2}, \vec{r}_\textrm{T})\right)\right],
\end{eqnarray*}
where $\vec{p}_\textrm{T}^{\,j_1}$ and $\vec{p}_\textrm{T}^{\,j_2}$
are \pt of two \textit{pseudo-jets}, sets of jets formed to maximize
their invariant mass, and $M_\textrm{T}$ is the transverse mass, i.e.,
$M_\textrm{T}(\vec{p}_\textrm{T}, \vec{q}_\textrm{T}) =
\sqrt{2\left(|\vec{p}_\textrm{T}||\vec{q}_\textrm{T}| -
  \vec{p}_\textrm{T}\cdot\vec{q}_\textrm{T}\right)}$. In words, among
the variations of two components into which \VEtmiss can be split,
\MTtwo is the minimum value of the larger value of the two transverse
masses which are formed from two pairs of a pseudo-jet and one of the
two \VEtmiss components. As can be seen in
Fig.~\ref{fig:CMS-SUS-15-003_Figure_001}, while \MTtwo is small for
QCD multijet events, it can be large for signal events.

In contrast to the Run~1 analyses
\cite{Khachatryan:2015vra,Chatrchyan:2012jx}, the search region in
this analysis is extended to include events with one jet
(\textit{monojet}). Therefore, the events in the search region have
one or more jets and no isolated electron, muon, or track. Because of
the conditions of the online triggers used in the analysis, the events
in the search region satisfies either $\HT > 200\GeV$ and $\MET >
200\GeV$ or $\HT > 1000\GeV$ and $\MET > 30\GeV$. Further, the events
with two or more jets are required to have $\MTtwo > 200\GeV$, $\dphi
> 0.3$ with \MET for the four largest \pt jets, and $|\MET-\MHT|/\MET
< 0.5$.

\begin{figure*}[!t]
  \centering
  \includegraphics[scale=0.5]{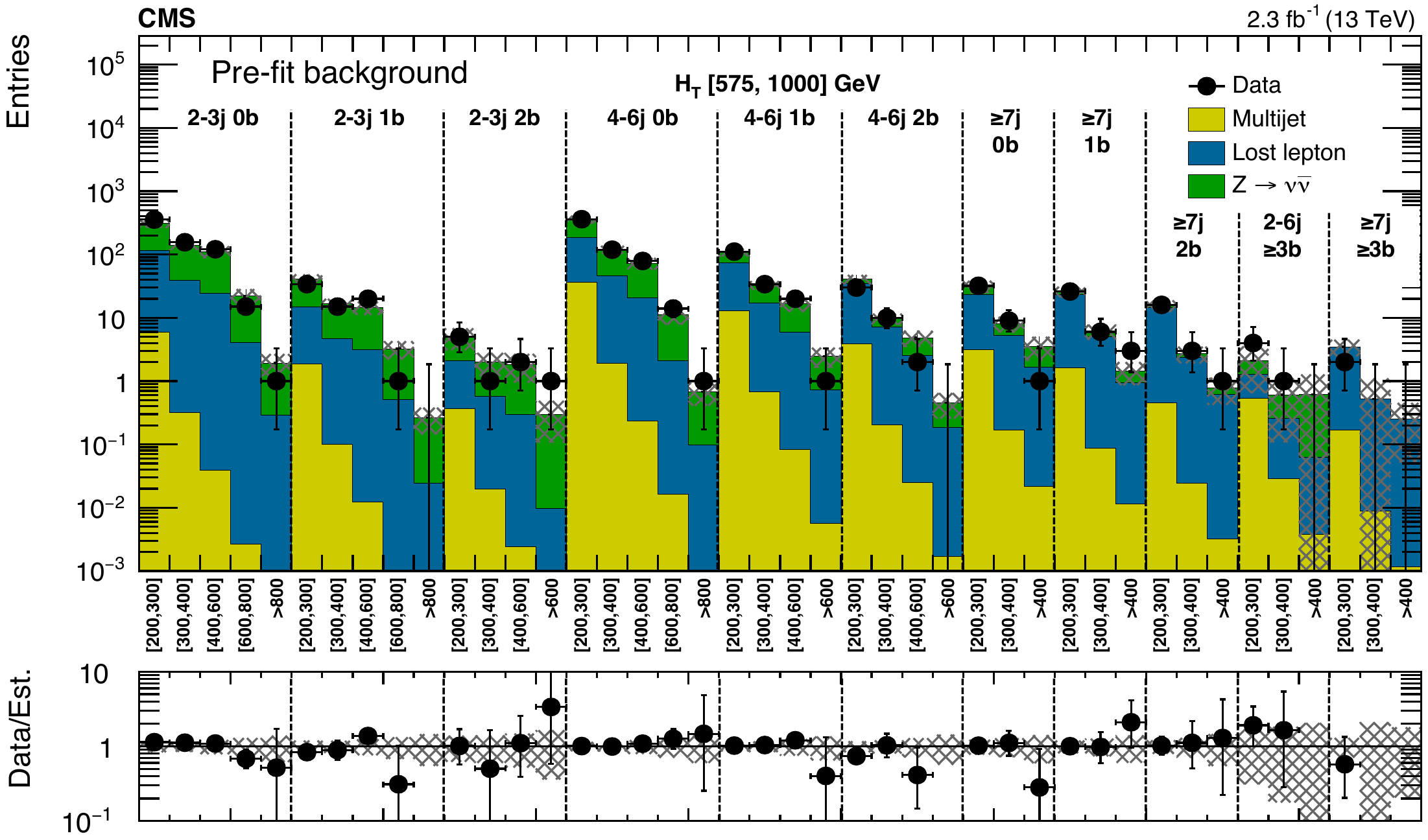}
  \caption{The observed number of the events and the SM background
    predictions in the \njet, \nb, and \MTtwo bins in one of the \HT
    bins in the search region of the analysis in
    Ref.~\cite{Khachatryan:2016xvy}.}
  \label{fig:CMS-SUS-15-003_Figure_008-b}
\end{figure*}

The events in the search region are categorized in the five \HT bins
and eleven \njet and \nb bins. The events with two ore more jets are
further categorized in the bins of \MTtwo. The boundaries of the
\MTtwo bins depend on the \HT bin and are illustrated on the top in
Fig.~\ref{fig:CMS-SUS-15-003_Figure-aux_003-4}. The monojet events are
categorized in the bins of the jet \pt, as shown on the bottom in
Fig.~\ref{fig:CMS-SUS-15-003_Figure-aux_003-4}. The analysis total has
172 categories.

The SM background is predicted for the QCD multijet, lost-lepton, and
\znnjets events. The data are consistent with the predictions.
Figure~\ref{fig:CMS-SUS-15-003_Figure_008-b} shows the comparison in
the \njet, \nb, and \MTtwo bins in one of the \HT bins. Similar
figures for the other bins are shown in
Ref.~\cite{Khachatryan:2016xvy}. In addition,
Ref.~\cite{Khachatryan:2016xvy} provides the numerical values of the
the predictions and observations in 14 aggravated bins for simpler
reinterpretations. Section~\ref{sec:interpretation} includes a summary
of the interpretations of the results.

\section{Search with razor variables}

\begin{wrapfigure}[14]{r}{7cm}
\centering
\hspace*{0.1cm}
\includegraphics[scale=0.4]{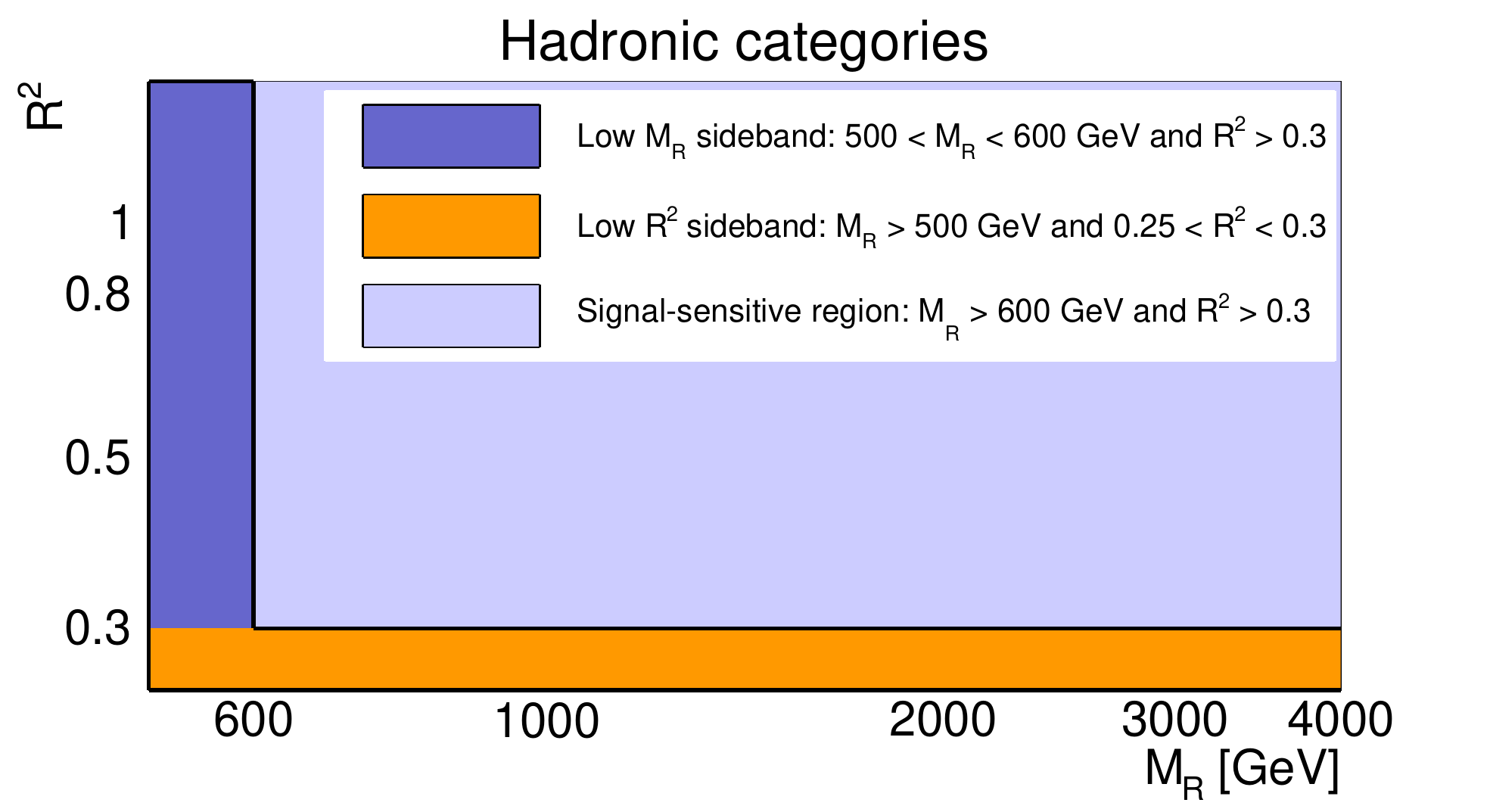}
\caption{The sidebands and the signal region on the
  $M_\textrm{T}^\textrm{R}$-$R^2$ plane in the multijet category in
  Ref.~\cite{CMS-PAS-SUS-15-004}}
\label{fig:CMS-PAS-SUS-15-004_Figure_002-a}
\end{wrapfigure}

The analysis reported in Ref.~\cite{CMS-PAS-SUS-15-004} uses the razor
variables in the definition of the search region and the background
prediction. The razor variables were used in SUSY searches in Run~1
\cite{Khachatryan:2016zcu,Khachatryan:2015exa,Khachatryan:2015pwa,Chatrchyan:2014goa,Chatrchyan:2012uea,Chatrchyan:2011ek}.
The analysis searches events with zero or one lepton, and four or more
jets.

In defining the razor variables, longitudinal momenta are used to find
an approximate center-of-mass frame of the patron-level process
\cite{Rogan:2010kb}. For events in SUSY models, the distribution of
the razor variable $M_\textrm{T}^\textrm{R}$, with the dimension of
mass, has a broad peak, and the distribution of the dimensionless
razor variable $R^2$ has a long tail. In contrast, for the events in
the SM processes, the distributions of both variables exponentially
decrease, which is used to model the background distribution on the
$M_\textrm{T}^\textrm{R}$-$R^2$ plane.

The analysis defines three event categories in the search region based
on the presence of an electron or muon. The events in the \textit{muon
  multijet} (\textit{electron multijet}) category are required to have
one muon with \pt > 20\GeV (one electron with \pt > 25\GeV),
$M_\textrm{T}$ > 120\GeV, 4 jets with \pt > 40\GeV,
$M_\textrm{T}^\textrm{R}$ > 400\GeV, and $R^2$ > 0.15. The events in
the \textit{multijet} category has 4 jets with \pt > 40\GeV or 2 jets
with \pt > 80\GeV, $M_\textrm{T}^\textrm{R}$ > 500\GeV, and $R^2$ >
0.25. The events in the muon multijet and electron multijet categories
(multijet category) are further categorized in 56 (35) bins of
$M_\textrm{T}^\textrm{R}$ and $R^2$ and 4 \nb bins.

A function of $M_\textrm{T}^\textrm{R}$ and $R^2$ with two additional
parameters is used to model the dependency of the distributions of the
SM background events on $M_\textrm{T}^\textrm{R}$ and $R^2$. This
function, which is fit in each \nb bin to the data in sidebands of the
signal region in $M_\textrm{T}^\textrm{R}$ and $R^2$, models the
distributions of the background events in the bins of
$M_\textrm{T}^\textrm{R}$ and $R^2$. The sidebands on the
$M_\textrm{T}^\textrm{R}$-$R^2$ plane are illustrated in
Fig.~\ref{fig:CMS-PAS-SUS-15-004_Figure_002-a}. No significant
deviations in the data from the prediction is observed. Figures
showing the comparisons of the data and the predictions for each bin
are provided in Ref.~\cite{CMS-PAS-SUS-15-004}. A summary of the
interpretations is given in Section~\ref{sec:interpretation}.

\section{Search with \alphat}

The analysis reported in Ref.~\cite{CMS-PAS-SUS-15-005} uses the
variables \alphat and \bdphimin to suppress the QCD multijet
background to a negligible level. These variables were used in SUSY
searches in Run~1
\cite{Khachatryan:2016pxa,Chatrchyan:2013mys,Chatrchyan:2012wa,Chatrchyan:2011zy,Khachatryan:2011tk}.

\begin{wrapfigure}[16]{r}{7.0cm}
\centering
\hspace*{0.02cm}
\includegraphics[scale=0.37]{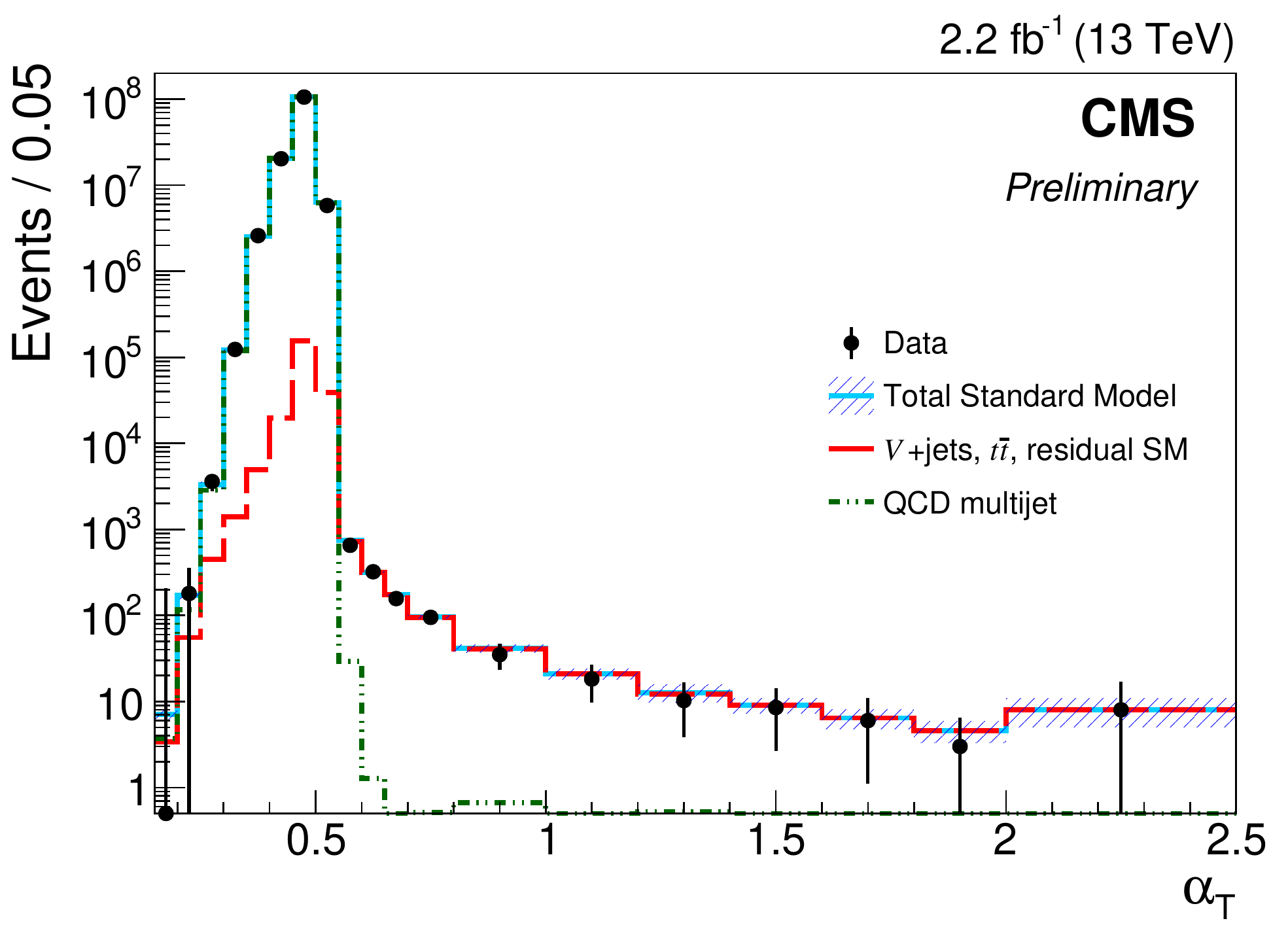}
\caption{The distribution of \alphat in the data and estimates of
  the contributions from the QCD multijet events and the other SM
  events \cite{CMS-PAS-SUS-15-005}.}
\label{fig:CMS-PAS-SUS-15-005_Figure_002-a}
\end{wrapfigure}

The \alphat variable \cite{Khachatryan:2011tk} is defined as
\begin{eqnarray*}
  \alphat = \frac{\sum_{i\in\textrm{jets}} \ET^{i} - \Delta \ET}{2\sqrt{\left(\sum_{i\in\textrm{jets}} \ET^{i}\right)^2 - {\HTm}^2}},
\end{eqnarray*}
where $\ET^i(\equiv E^i\sin\theta^i)$ is the jet transverse energy,
and $\Delta\ET$ is the difference in \ET of the two
\textit{pseudo-jets}, two sets of jets into which the jets in the
event are combined so as to minimize $\Delta\ET$. As can be seen in
Fig.~\ref{fig:CMS-PAS-SUS-15-005_Figure_002-a}, while the value of
\alphat for QCD multijet events do not become much larger than 0.5, it
can be for events from the other processes. The \bdphimin is the
minimum of the azimuthal angle between a jet and \MHT recomputed
without the jet, i.e.,
\begin{eqnarray*}
  \bdphimin =
  \min_{i\in\textrm{jets}}\Delta\phi(\vec{p}_{\textrm{T}i},
  \vec{H}_\textrm{T}^{\textrm{miss}} + \vec{p}_{\textrm{T}i}).
\end{eqnarray*}

The events in the search region do not have an isolated electron,
muon, or track and satisfy the conditions that $\njet \ge 1$, the
largest jet $\pt > 100\GeV$, $\HT > 200\GeV$, $\MHT > 130\GeV$,
$\MHT/\MET < 1.25$ and $\bdphimin > 0.5$. In addition, the events with
$\HT < 800\GeV$ have $\alphat$ greater than a certain threshold
between 0.52 and 0.65 depending on \HT.

In the Run~1 analyses
\cite{Khachatryan:2016pxa,Chatrchyan:2013mys,Chatrchyan:2012wa,Chatrchyan:2011zy,Khachatryan:2011tk},
the events in the search region are generally required to have at
least two jets with \pt greater than 100\GeV. The analysis in
Ref.~\cite{CMS-PAS-SUS-15-005}, on the other hand, categorizes the
events that satisfy this requirement as \textit{symmetric} and
includes two additional categories: \textit{asymmetric} and
\textit{monojet}. An event in the asymmetric category have one jet
with \pt greater than 100\GeV and one or more jets with \pt between
40\GeV and 100\GeV. An event in the monojet category has one jets with
\pt greater than 100\GeV and no 2$^{\textnormal{nd}}$ jet with \pt
greater than 40\GeV. The events in each of these three categories are
further categorized in the bins of \njet, \nb, \HT, and \MHT.
 
The SM background is predicted from the single-muon, double-muon, and
single-photon CRs. The observed numbers of the events and the
predictions, provided in each bin of \njet, \nb, and \HT in tables in
Ref.~\cite{CMS-PAS-SUS-15-005}, are consistent. A summary of the
interpretations is shown in the next section.

\section{Interpretation}
\label{sec:interpretation}

\begin{figure*}[!b]
  \centering
  \includegraphics[scale=0.65]{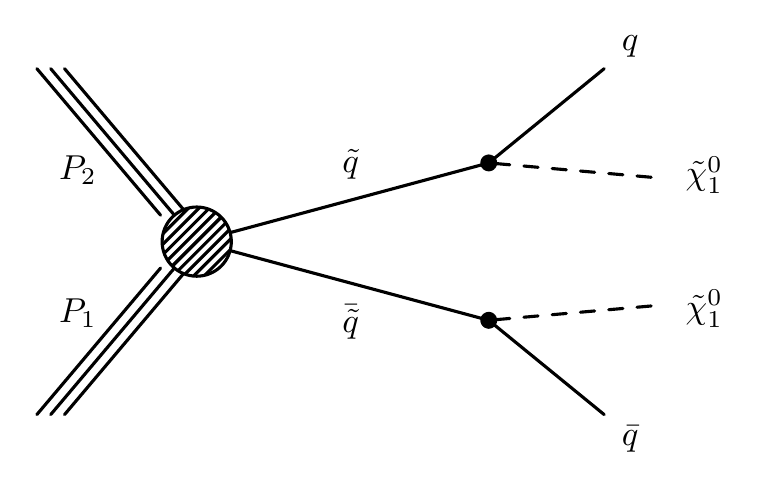}
  \includegraphics[scale=0.65]{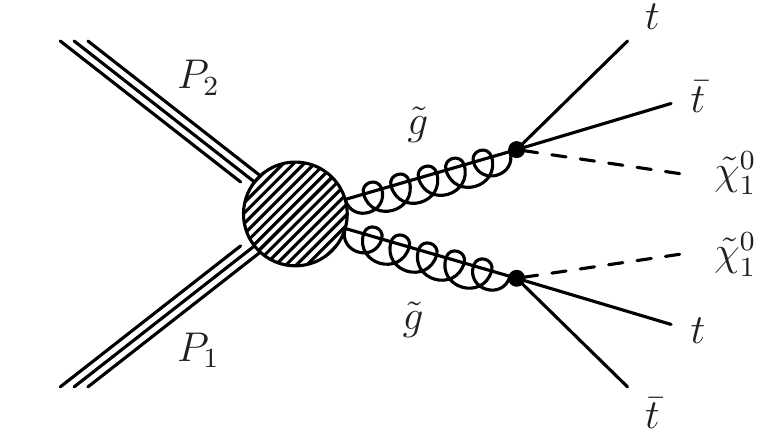}
  \includegraphics[scale=0.65]{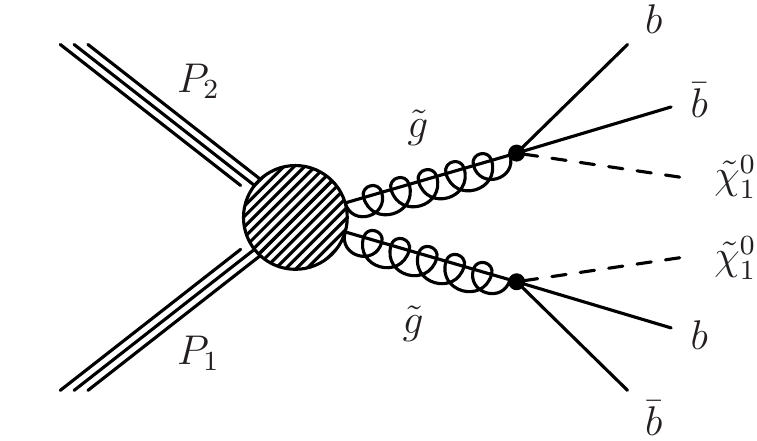}
  \includegraphics[scale=0.65]{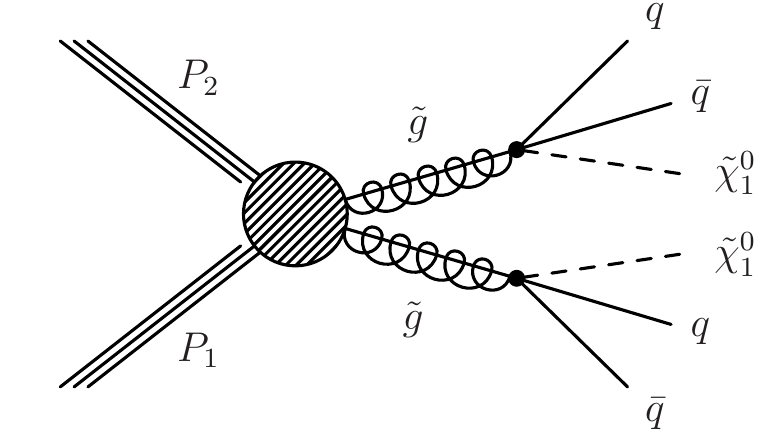}
  \caption{Diagrams for the gluino pair and squark pair productions in
    simplified models: T2qq \textit{(top left)}, T1tttt \textit{(top
      right)}, T1bbbb \textit{(bottom left)}, T1qqqq \textit{(bottom
      right)}.}
  \label{fig:sms_diagrams}
\end{figure*}

This section summarizes interpretations of the results in the squark
pair and gluino pair productions in \textit{simplified models} for
which diagrams are show in Fig~\ref{fig:sms_diagrams}, i.e., T2qq,
T1tttt, T1bbbb, and T1qqqq.
Refs.~\cite{Khachatryan:2016kdk,Khachatryan:2016xvy,CMS-PAS-SUS-15-004}
include interpretations in other simplified models as well.

\begin{figure}[!h]
  \begin{minipage}[t]{0.45\linewidth}
    \centering

    \includegraphics[scale=0.35]{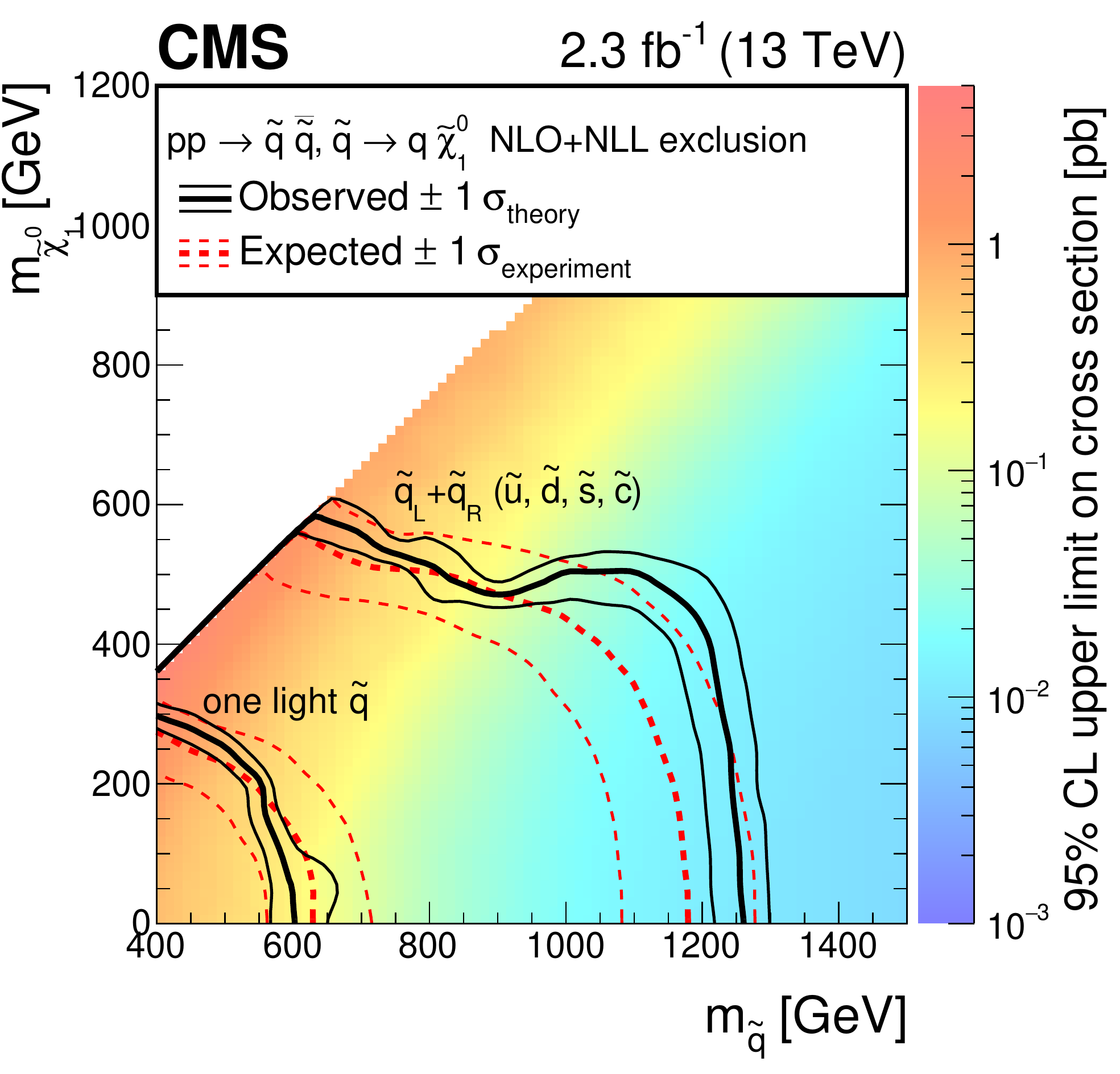}
    \caption{The limit on the production cross section of squark pairs
      and on the masses on the \PSGcz-\PSQ mass plane in T2qq for the
      case in which both chiral states of all light-flavored squark
      have the same mass and the case in which one chiral state of one
      light-flavored squark enters the model
      \cite{Khachatryan:2016xvy}.}
    \label{fig:CMS-SUS-15-003_Figure_011-c}

  \end{minipage}
  \hspace{0.5cm}
  \begin{minipage}[t]{0.45\linewidth}
    \centering

    \includegraphics[scale=0.35]{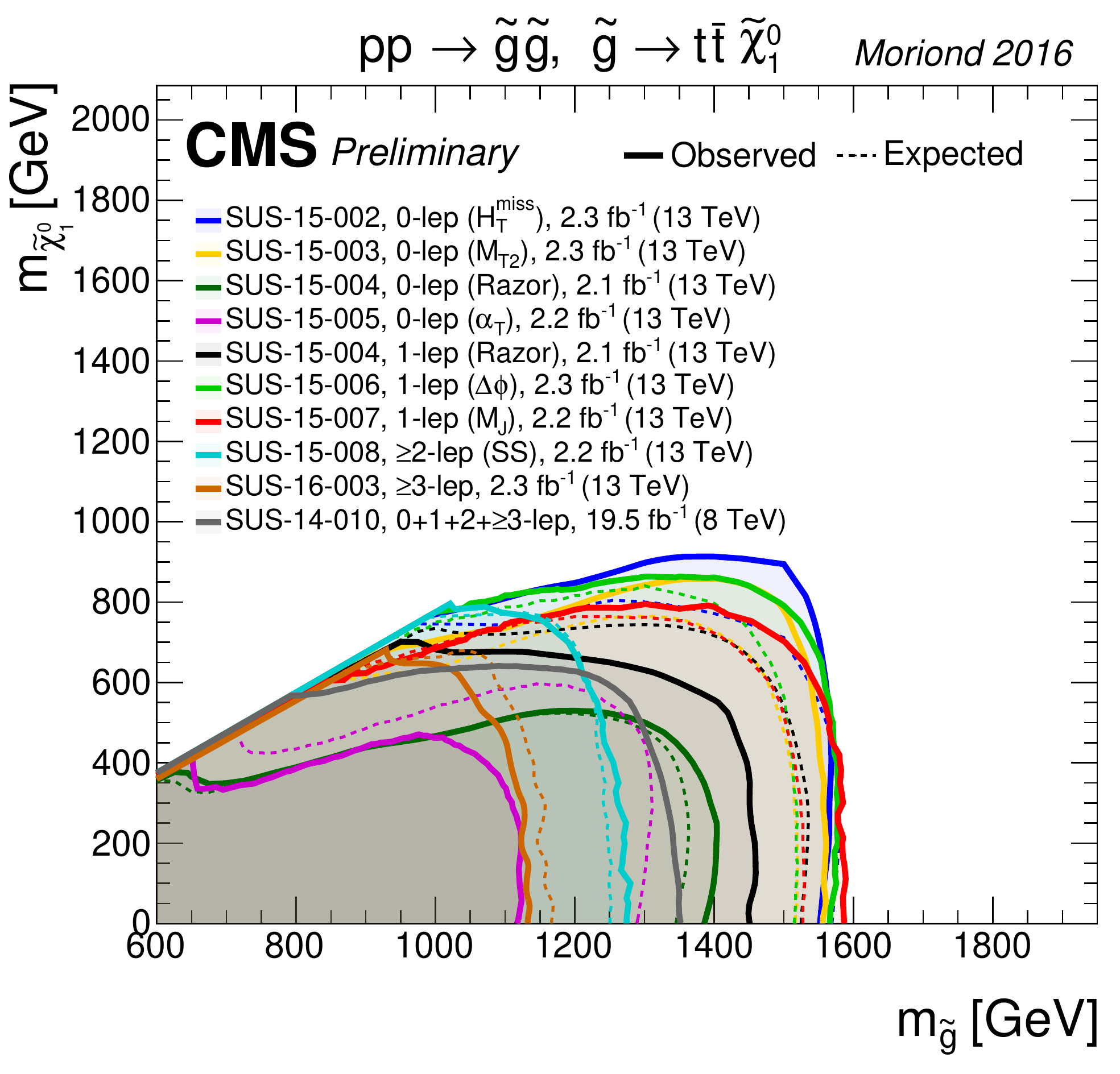}
    \caption{The limits on the \PSGcz-\PSg mass plane for
      T1tttt placed by the four analyses summarized in these
      proceedings
      \cite{Khachatryan:2016kdk,Khachatryan:2016xvy,CMS-PAS-SUS-15-004,CMS-PAS-SUS-15-005,CMS-PAS-SUS-16-004},
      four other analyses with the data collected during the same
      period
      \cite{CMS-PAS-SUS-15-006,Khachatryan:2016uwr,Khachatryan:2016kod,CMS-PAS-SUS-16-003},
      and one analysis with Run~1 data \cite{CMS:2014dpa}. The solid
      lines show the observed limits, and dashed lines show the
      expected limits.}
    \label{fig:T1tttt_Moriond}

  \end{minipage}

  \begin{minipage}[t]{0.45\linewidth}
    \centering

    \includegraphics[scale=0.35]{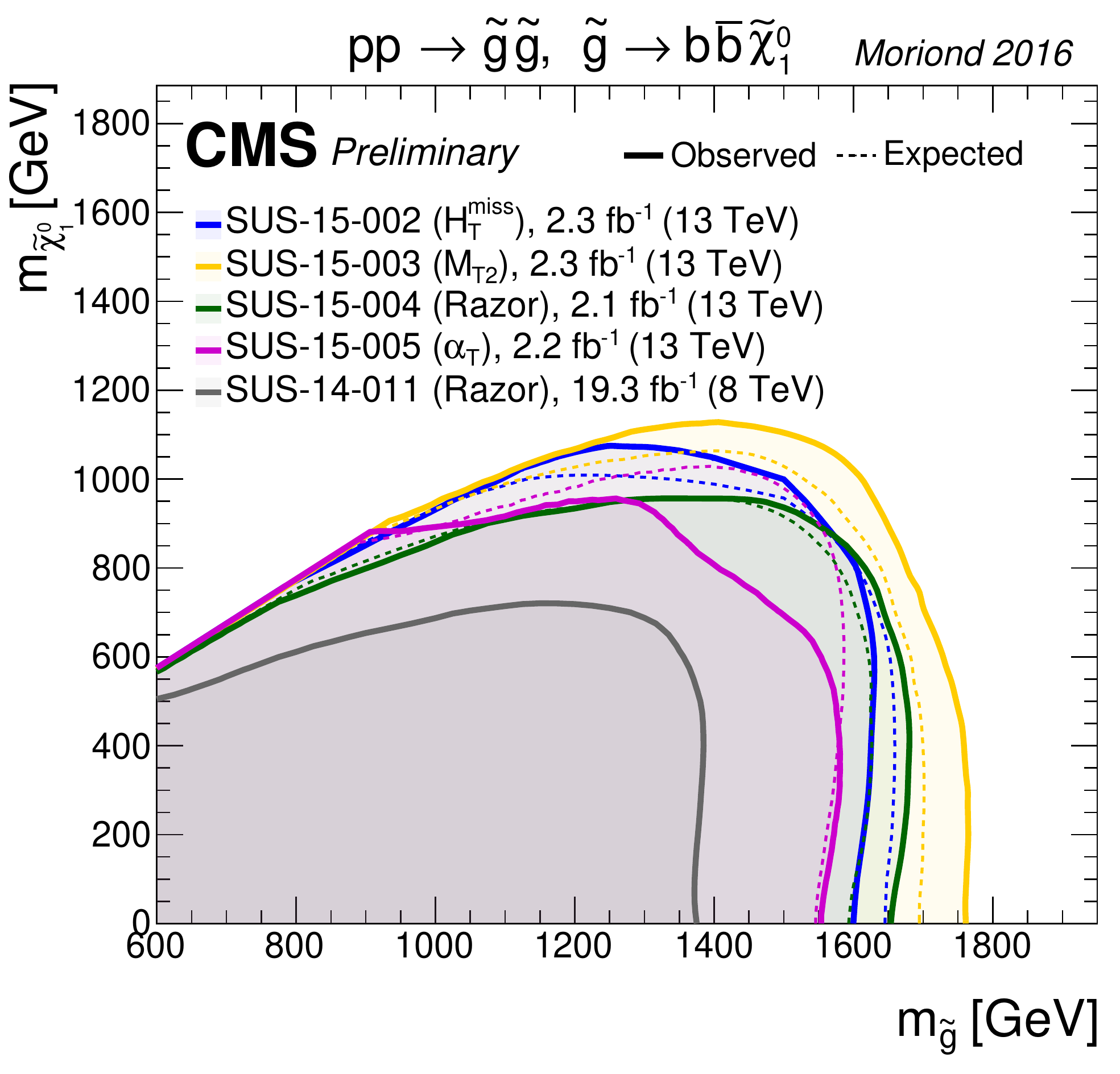}
    \caption{The limits on the \PSGcz-\PSg mass
      plane for T1bbbb placed by the four analyses summarized in these
      proceedings
      \cite{Khachatryan:2016kdk,Khachatryan:2016xvy,CMS-PAS-SUS-15-004,CMS-PAS-SUS-15-005}
      and one analysis with Run~1 data \cite{CMS-PAS-SUS-14-011}.}
    \label{fig:T1bbbb_Moriond}

  \end{minipage}
  \hspace{0.5cm}
  \begin{minipage}[t]{0.45\linewidth}
    \centering

    \includegraphics[scale=0.35]{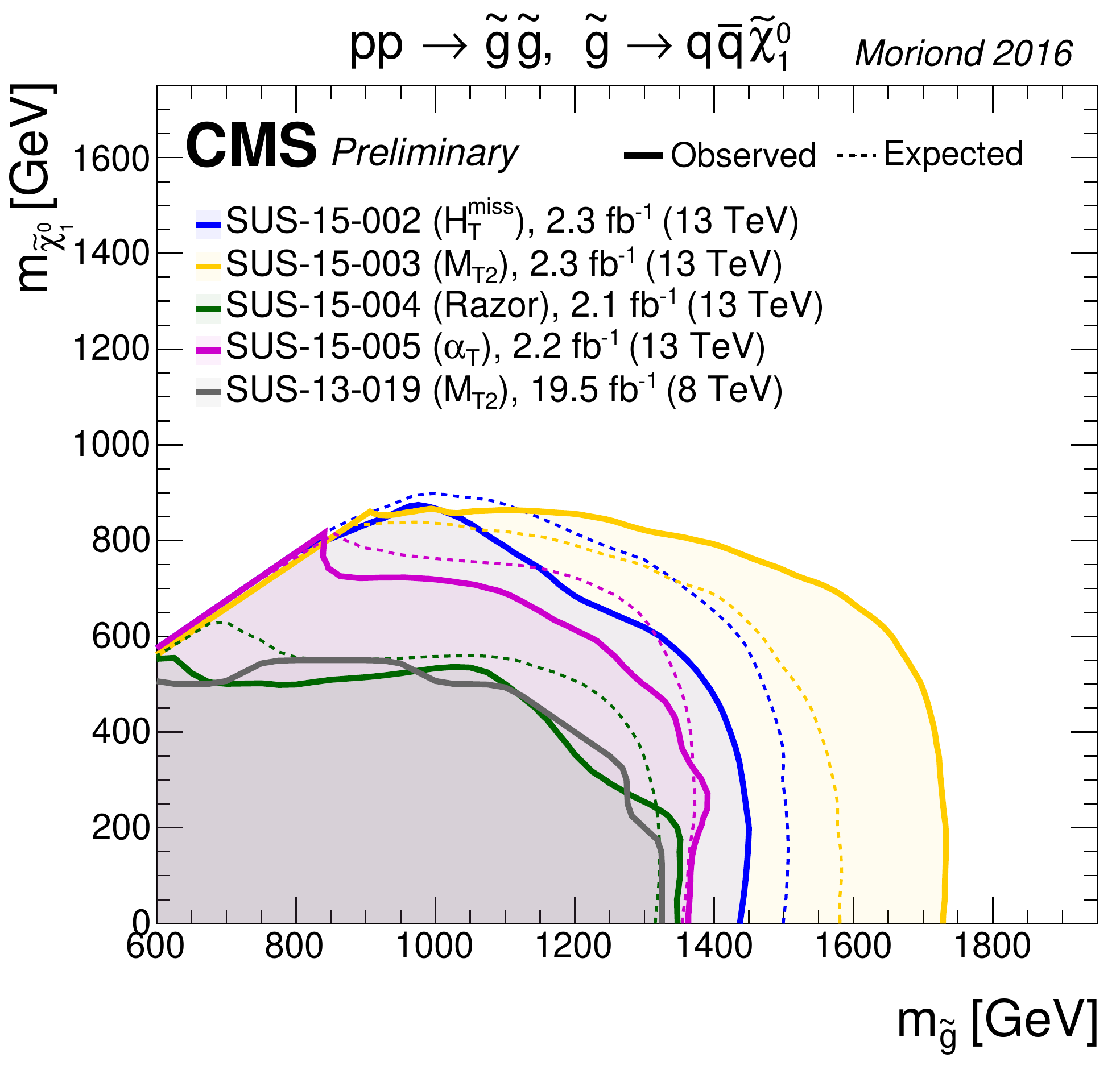}
    \caption{The limits on the \PSGcz-\PSg mass plane for
      T1qqqq placed by the four analyses summarized in these
      proceedings
      \cite{Khachatryan:2016kdk,Khachatryan:2016xvy,CMS-PAS-SUS-15-004,CMS-PAS-SUS-15-005}
      and one analysis with Run~1 data \cite{Khachatryan:2015vra}.}
    \label{fig:T1qqqq_Moriond}

  \end{minipage}
\end{figure}

Figure~\ref{fig:CMS-SUS-15-003_Figure_011-c} shows an interpretation
of the results of the analysis summarized in Section~\ref{sec:mt2} in
T2qq. On the neutralino-squark mass plane, the upper limits on the
production cross sections are shown as the color scale. The thick
lines show the observed lower limits on the masses of neutralinos and
squarks. The limits are shown for two cases: that both chiral states
of all light-flavored squark have the same mass and that one chiral
state of one light-flavored squark enters the model. The thick dashed
lines show the expected lower mass limits, the limits that would be
obtained if the predicted number of the events were actually observed.
For the former case, the squark mass up to about 1260\GeV and the
neutralino mass up to about 580\GeV are excluded. For the latter case,
the squark and neutralino masses up to 600\GeV and 300\GeV are
excluded.

Figures~\ref{fig:T1tttt_Moriond}, \ref{fig:T1bbbb_Moriond}, and
\ref{fig:T1qqqq_Moriond} summarize observed and expected lower mass
limits on the neutralino-gluino mass planes in T1tttt, T1bbbb, and
T1qqqq, respectively, placed by the analysis summarized in these
proceedings. Figures similar to
Fig.~\ref{fig:CMS-SUS-15-003_Figure_011-c}, i.e., showing the cross
section limits as the color scale, for these models are provided in
Refs.~\cite{Khachatryan:2016kdk,Khachatryan:2016xvy,CMS-PAS-SUS-15-004,CMS-PAS-SUS-15-005,CMS-PAS-SUS-16-004}.
Figure~\ref{fig:T1tttt_Moriond} includes the limits placed by other
analyses with the data collected during the same period
\cite{CMS-PAS-SUS-15-006,Khachatryan:2016uwr,Khachatryan:2016kod,CMS-PAS-SUS-16-003}.
In each figure, a limit placed by an analysis with the 19 fb$^{-1}$ of
data at 8\TeV collected during Run~1 is also shown for a comparison
\cite{CMS:2014dpa,CMS-PAS-SUS-14-011,Khachatryan:2015vra}. The gluino
mass up to 1750\GeV in the limit of massless neutralinos and the
neutralino mass up to 1125\GeV in the case of the light gluinos are
excluded. These limits are significantly raised from the Run~1
results.

\clearpage

\section{Summary}

We have searched for pair productions of squarks and gluinos in
hadronic channels in 2.3~fb$^{-1}$ of proton-proton collisions at
$\sqrt{s}=13$~TeV collected with the CMS detector in 2015, the first
year of LHC Run~2. The number of the event that we observed in the
search region of each analysis was consistent with the standard model
prediction. The limits on the production cross sections and masses of
squarks and gluinos in simplified models have been significantly
extended from the previous results. The data collection with the CMS
detector in 2016 has been successful. We have released updates on the
searches summarized in these proceedings with 12.9~fb$^{-1}$ of data
collected in 2016
\cite{CMS-PAS-SUS-16-014,CMS-PAS-SUS-16-015,CMS-PAS-SUS-16-016}. At
the time of this writing, 28.8~fb$^{-1}$ of data have been collected.
We are analyzing these data and continue to test the predictions of
supersymmetric extensions of the standard models.

\bibliographystyle{JHEPm}

\begingroup
    \small
    \setlength{\bibsep}{0pt}
    \setstretch{0.85}
    \bibliography{proceedings}
\endgroup

\end{document}